\documentclass[%
 reprint,
 amsmath,amssymb,
 aps,
]{revtex4-1}

\usepackage{color}
\usepackage{graphicx}
\usepackage{dcolumn}
\usepackage{bm}

\usepackage{color}

\begin{document}
\preprint{APS/123-QED}
\title{Rydberg level shift due to the electric field generated by Rydberg atom collision induced ionization in cesium atomic ensemble}

\author{Xin Wang${}^{1}$}
 \altaffiliation{These two authors contributed equally to this work}
\author{Jun He${}^{12*}$} 
 \email{hejun@sxu.edu.cn}
\author {Jiandong Bai${}^{1}$}
\author{Junmin Wang${}^{12}$}
 \email{wwjjmm@sxu.edu.cn}
\affiliation{%
{${}^{1}$State Key Laboratory of Quantum Optics and Quantum Optics Devices, and Institute of Opto-Electronics, Shanxi University, Tai Yuan 030006,  Shanxi Province, People's Republic of China}\\{${}^{2}$Collaborative Innovation Center of Extreme Optics of the Ministry of Education and Shanxi Province, Shanxi University, Tai Yuan 030006, Shanxi Province, People's Republic of China}
}%

\begin{abstract}
We experimentally studied the Rydberg level shift caused by the electric field, which is generated by Rydberg atom collision induced ionization in a cesium atomic ensemble. The density of charged particles caused by collisions between Rydberg atoms is changed by controlling the ground-state atomic density and optical excitation process. We measured the Rydberg level shift using Rydberg electromagnetically-induced-transparency (EIT) spectroscopy, and interpreted the physical origin using a semi-classical model. The experimental results are in good agreement with the numerical simulation. These energy shifts are important for the self-calibrated sensing of microwave field by the employing of Rydberg EIT. Moreover, in contrast to the resonant excitation case, narrow-linewidth spectroscopy with high signal-to-noise ratio would be useful for high-precision measurements.
\end{abstract}


\maketitle


\section{Introduction}
Recent progress in technical control has offered possibilities for studying the interactions between pure quantum mechanical systems and their coupling to reservoirs, which can be extended as unique tools for quantum sensing \cite{ref-journal1}. Alkali metal atoms are convenient, in~terms of the technical requirements needed for sensors. Neutral atoms in chip-scale atomic vapor have been utilized as a device for many applications, including atomic clocks \cite{ref-journal2}, atomic magnetometers cell \cite{ref-journal3}, and for radio-frequency (RF) fields sensors \cite{ref-journal4}. Atom-based electromagnetic field sensor technologies hold great promise for realizing such applications. The~most advanced experimental demonstrations include high dynamic field ranges (exceeding 120 dB), high-intensity RF fields (up to 10 kV/m) \cite{ref-journal5}, broadband RF detection (on the order of THz) \cite{ref-journal6}, and~high sensitivity ($up\mu V\cdot$ cm$^{-1}$/Hz$^{1/2}$) \cite{ref-journal4,ref-journal7}. Recent advances have experimentally achieved sensitivities of 55~nV $\cdot$ cm$^{-1}$/ Hz$^{1/2}$, which approaches the quantum projection noise limit \cite{ref-journal8}; however, going beyond this, it is possible by using squeezed states or Schrodinger-cat states of light field \cite{ref-journal9}.

Recent advances have provided new capabilities in RF or microwave (MW) sensing. Typical~high-frequency microwave fields have a wavelength range on the order of millimeters or centimeters. In~chip-scale atomic vapor cell, the~signal is limited by atomic density. In~contrast to the characteristics of atomic spin, Rydberg atoms with high principal quantum number have exaggerated characteristics including dipole--dipole or Van der Waals interactions \cite{ref-journal10}. Using a buffer-gas-filled cell to improve atomic coherence is not feasible, as~collision will destroy the electric dipole coherence. The~high atomic density can be obtained by ultraviolet light desorption effect or heating the atomic vapor cell. Under~high-density atomic gas, the~sensing accuracy and sensitivity are limited by the spectral broadening \cite{ref-journal11}, including Doppler broadening, collision broadening, and~interaction dephasing induced by Rydberg atom collisions. The~dephasing observed in the experiments has been explained by averaging the effects resulting from Van den Waals interaction of Rydberg--Rydberg interaction or coupling between Rydberg atoms and the inner surface of the glass cell \cite{ref-journal12,ref-journal13,ref-journal14}.

In this paper, by~measuring the Rydberg electromagnetically-induced-transparency (EIT) signal, we observed Rydberg level shift caused by Stark interaction of Rydberg atoms and the mean electric field which is generated by Rydberg atom collision induced ionization in a cesium atomic ensemble. The~results are obtained by means of Doppler-free spectroscopy with high signal-to-noise ratio, which~is essential for high-precision measurements. Understanding the microscopic mechanism of charged particles production in thermal Rydberg atomic vapor and its corresponding electric field distribution is an important part of future experiments and~applications. 
 
\section{Setup and measurement}

\begin{figure}[htbp]
\centering
\includegraphics[width=8.5cm]{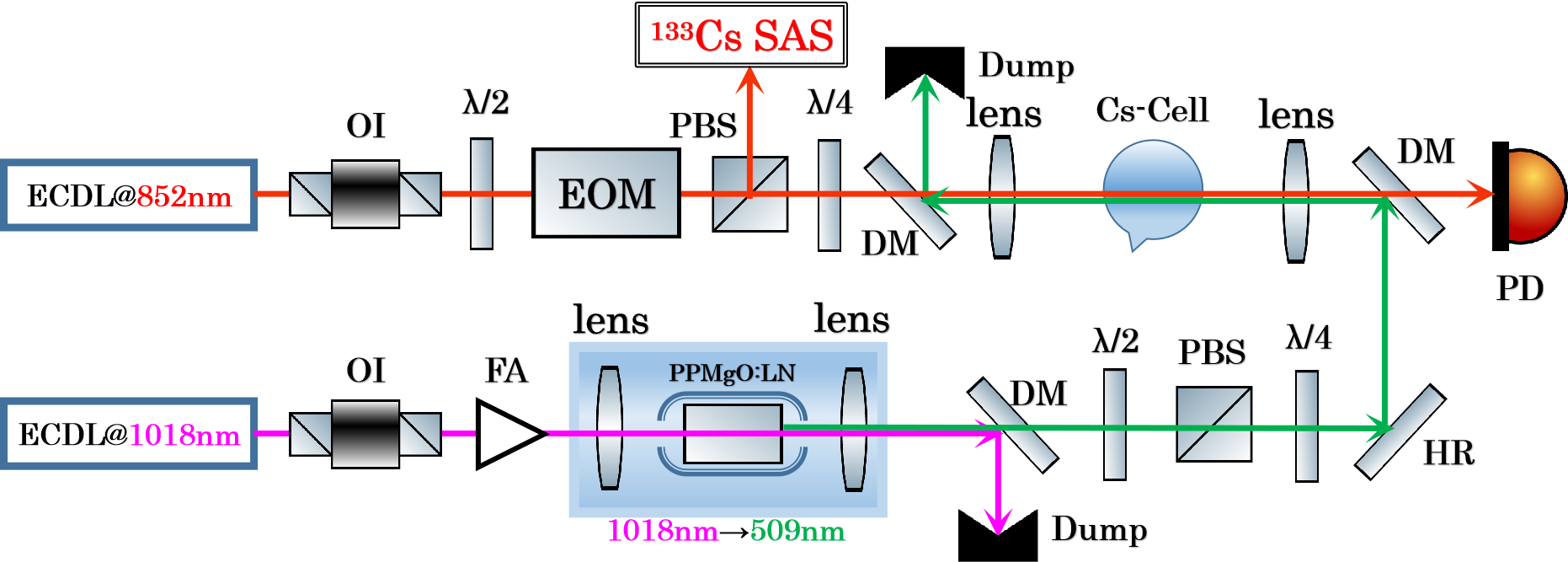}
\caption{Schematic diagram of the experimental setup.{{An ECDL at 852 nm is used as a probe laser, and an~ECDL at 1018 nm is frequency-doubled and used as a 509-nm coupling laser. The~frequency-doubling crystal is a PPLN. A~SAS (saturation absorption spectroscopy) setup is used as a frequency reference for the hyperfine structure of the Cs D2 line,}} where OI is optical isolator; EOM, electro-optic modulator; $\lambda$/2, half-wave plate; $\lambda$/4, quarter-wave plate; PBS, polarization beam splitter cube; DM, dichroic mirror;  FA, fiber~amplifier; PD, photodiode (New Focus, model 2051); HR, high reflection mirror; SAS, saturation absorption~spectroscopy. }
\end{figure}

\begin{figure}[htbp]
\centering
\includegraphics[width=9cm]{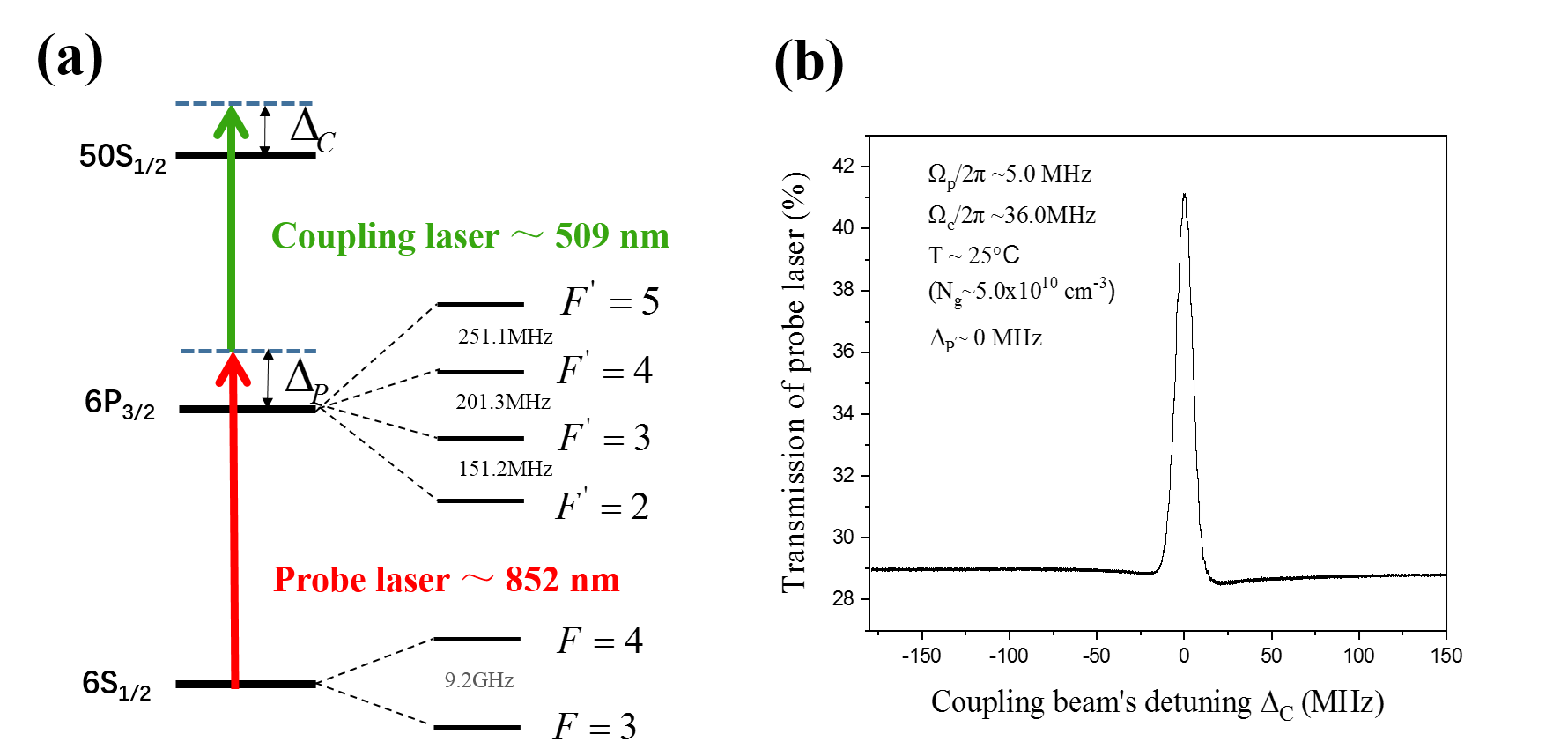}
\caption{(\textbf{a}) Energy level schematic of the ladder-type EIT of Cs atoms with Rydberg state (50S$_{1/2}$). $\Delta_P$ is the detuning of probe laser, $\Delta_C$ is the detuning of coupling laser; (\textbf{b}) Typical Rydberg EIT signal. The~852-nm probe laser is detuned by $\Delta_P$ from the Cs $6S_{1/2}(F=4)\rightarrow 6P_{3/2}(F'=5)$ transition (here, $\Delta_P=0$), while the 509-nm coupling laser is scanned across the Cs $6P_{3/2}(F'=5)\rightarrow 50S_{1/2}$ Rydberg transition.
}
\end{figure}

The experimental setup is shown in Figure 1. An~external-cavity diode laser (ECDL) at 852 nm is used as a probe laser with a typical linewidth of $\sim $90 kHz. The~optical power of a 1018-nm ECDL is amplified to $\sim$5 W by a {{fiber amplifier(FA), and~then frequency-doubled by a single-passed periodically poled lithium niobate (PPLN) crystal to 509 nm.}} The two beams are overlapped in cesium (Cs) atomic vapor in a counter-propagating configuration. The~spherical cesium cell has a radius of $\sim $1.2~cm to match the Rayleigh length of the focused beams ($\sim$40\,$\mu$m waist radius). We first performed the saturation absorption spectroscopy and determined the zero detuning of the 852-nm probe laser. {{The~wavelength meter (HighFinesse WS-7, with~wavelength deviation sensitivity of about 10~MHz) was calibrated using the Cs $6S_{1/2}(F=4)\rightarrow 6P_{3/2}(F'=5)$ hyperfine transition line. With~the calibrated wavelength meter, we measured the wavelength of probe beam (when set to a certain detuning) and the coupling beam.}} The ladder-type Rydberg EIT signal is observed by scanning the frequency of 509-nm coupling laser while the 852-nm probe laser's frequency is fixed. The~background-free EIT spectrum with high signal-to-noise ratio {{can be regarded as a low-noise detection system,}} which decreases the intensity noise and phase noise. Figure 2 shows the relevant energy level schematic and the transmitted intensity of a typical Rydberg EIT signal at 25 $^{\circ}$C. The~linewidth of the EIT signal is about 10 MHz, as~measured by a RF modulation~spectroscopy.

\begin{figure}[bp]
\centering
\includegraphics[width=8.5cm]{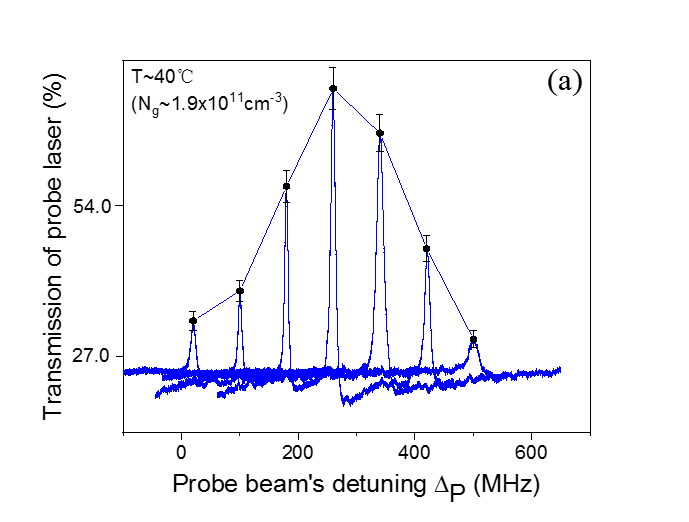}
\end{figure}
\begin{figure}[htbp]
{\includegraphics[width=8.5cm]{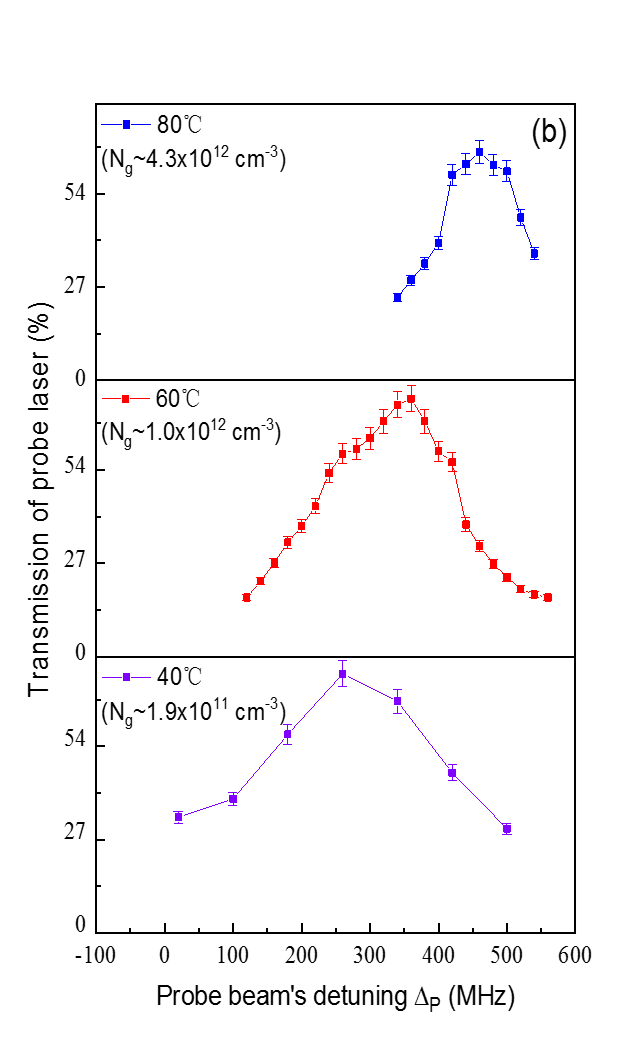}
\includegraphics[width=8.5cm]{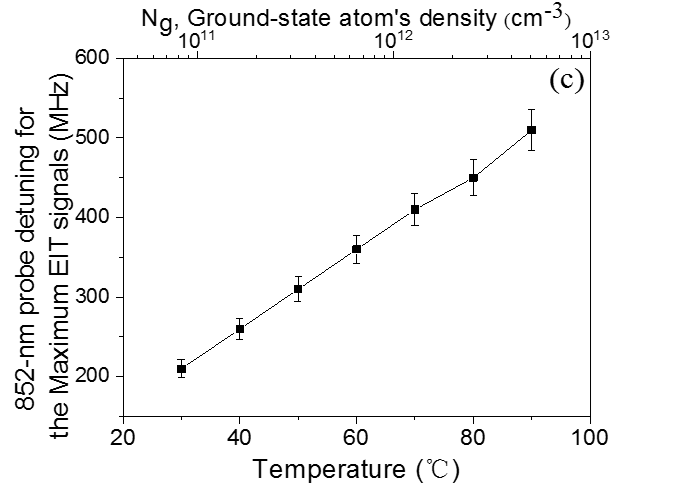}}
\caption{(\textbf{a}) At temperature of 40 $^{\circ}$C, the~Rydberg EIT signal is obtained by scanning the 509-nm coupling laser frequency with different 852-nm probe laser detuning; (\textbf{b}) Rydberg EIT signal intensity versus probe laser detuning for different atomic cell temperature; (\textbf{c}) The 852-nnm probe laser detuning of the maximum intensity EIT signal changes with temperature (ground-state atomic density), including the three temperature corresponding data points in (\textbf{b}). Rabi frequency of 852-nm probe laser is $\sim$5.0 MHz, and that of the 509-nm coupling laser is $\sim$36.0~MHz. }
\end{figure}

We study the relationship between the Rydberg EIT signal and the temperature as well as laser detuning in a ladder-type system. Figure 3(a) shows the dependence of the Rydberg EIT signal on probe detuning at a cell temperature of 40 $^{\circ}$C, where each small peak in the figure represents an EIT signal. Zero~detuning of the probe laser {{is identical with the}} Cs $6S_{1/2}(F=4)\rightarrow 6P_{3/2}(F'=5)\rightarrow 50S_{1/2}$ cascade resonance transition. We fix the 852-nm laser frequency {{at a chosen detuning}} and scan the 509-nm laser frequency {{in order to get EIT signal}}. When the detuning of the probe laser is $\sim$260~MHz, the~EIT signal’s strength is the largest, which is caused by the electric field generated by the collisions of Rydberg atoms causing the Rydberg level shift, a~detailed explanation can be found in the discussion part. In Figure 3(b), only the envelope curves are shown, for~40, 60, and 80 $^{\circ}$C. Figure 3(c) shows that the 852-nm laser detuning for the maximum intensity of EIT signals changed with temperature (ground-state atomic density). It can be seen that probe laser detuning with the maximum EIT signal intensity increased {{nearly linear}} with increasing temperature. For~example, at temperature of 90 $^{\circ}$C, the~probe laser detuning with the maximum EIT signal intensity is $\sim$510 MHz. Under~high temperature conditions, the~Rydberg EIT signal for the Cs $6S_{1/2}(F=4)\rightarrow 6P_{3/2}(F'=5)\rightarrow 50S_{1/2}$ cascade transition (zero detuning of the probe laser) is very weak. {{We observe that the linewidth of the EIT signal does not change significantly as the temperature increases, and is $\sim$10 MHz.}} There is no obvious linewidth broadening, which~means that the electric field-induced shift is not a dissipative~process.

\section{Analysis and discussions}

The most advanced experiments demonstrate that collisions in a Rydberg atomic ensemble will yield ionization which will produce positive ions and electrons \cite{ref-journal15}. We know that, under~the conditions of a weak electric field, the~Rydberg level shift will be dominated by Stark shift. The~dependence of the Rydberg level shift on the weak electric field can be measured by Rydberg EIT spectroscopy. We expand this idea and use a semi-classical model to explain the physical phenomena of the Rydberg level shift. In~a thermal Rydberg atomic ensemble, the~collisions between Rydberg atoms and the collisions between Rydberg atoms and ground-state atoms induce charged particles (positive ions and electrons), which cause avalanche ionization \cite{ref-journal15,ref-journal16,ref-journal17}. Here, the~photo-ionization process is very weak, it can be ignored. In~our system, the~Rydberg atomic density is obtained by the combined effect of the ground-state cesium atomic density and the two-step Rydberg excitation $R_{exc}$, under~the premise that the other experimental conditions (i.e., probe Rabi frequency and coupling Rabi frequency) remain unchanged, and~ionization leads to the reduction of the Rydberg atomic~density.

In the theoretical model, {{positive ions and electrons}} will generate an inhomogeneous local electric field in the local space of the atomic vapor, and~the local electric field interacting with Rydberg atoms will reduce the Rydberg energy and cause a Rydberg level shift. We assume the electron density and the positive ion density to be equal, and~both parameters are given by  $N_{C}$. The~dependence of charged particles'density and Rydberg atomic density can be obtained by solving the following rate equations \cite{ref-journal15}:

\begin{equation}
{\dot{N}_{Ryd}}=-{N_{Ryd}}{N_g}\sigma_g\overline{v}-{N_{Ryd}}{N_{Ryd}}\sigma_{Ryd}\overline{v}'+R_{exc}
\end{equation}
\begin{equation}
{\dot{N}_{C}}={N_{Ryd}}{N_g}\sigma_g\overline{v}+{N_{Ryd}}{N_{Ryd}}\sigma_{Ryd}\overline{v}'-{N_{C}}\Gamma_d
\end{equation}
where $N_{Ryd}$ and $N_{C}$ are Rydberg atom density and induced density of charged particles, respectively. $N_g$ is the ground-state atomic density; $\sigma_g$is cross-section for Rydberg-ground state atomic collisions, $\sigma_{Ryd}$ is geometric cross-section of Rydberg atomic collisions \cite{ref-journal18,ref-journal19}, $\overline{v}$ is the mean relative velocity of Rydberg atoms colliding with ground-state atoms, and~ $\overline{v}'$ is the mean relative velocity of Rydberg atoms colliding with other Rydberg atoms. The~expected value of $\sigma_g$ is 0.06$\sigma_{Ryd}$ \cite{ref-journal15}, with~$\sigma_{Ryd}=\pi(a_0^2n^{*4})$ \cite{ref-journal15,ref-journal20} (where $a_0$ is the Bohr radius) being the geometric cross-section of Rydberg state with an effective principal quantum number $n^*$. $R_{exc}= {\Omega_p^2}/({\Omega_p^2+\Omega_c^2})\cdot{N_g}{\Omega_p\Omega_c}/{\Delta}$ is the Rydberg component in superposition dark state of EIT, where $\Delta$ is two-photon detuning. Here, $\Gamma_d$ contains charged particles recombination and decay, and the~rate $\Gamma_d=\Gamma_t+2{N_g}\sqrt{8K_BT/(\pi m_e)}\sigma_r\rho_{c}$ represents the losses in the charged particle. $\Gamma_t$ is the charged particle diffusion rate. The~most obvious contribution is that $\Gamma_t$ describes the motional decay parameters of charged particles leaving the interaction volume due to thermal motion, which vary slightly with temperature. The value of $\Gamma_t$ is $\sim$0.25 MHz. The~second term in $\Gamma_d$ is represented by 2$\Gamma_r$, which describes the recombination of ions with electrons into neutral particles. Here, $K_B$ is the Boltzmann constant, $m_e$ is the electron mass, $\rho_{c}$ is the charged particles population, and we take $\rho_{c}\approx$ 1 in our model. The~expected value for $\sigma_r$ can be estimated \cite{ref-journal21} in a semi-classical treatment with Kramer’s formula. Compared with the ion diffusion rate of 0.25 MHz, 2$\Gamma_r$  is very small and can be~ignored.

\begin{figure}[bp]
\centering
\includegraphics[width=8.5cm]{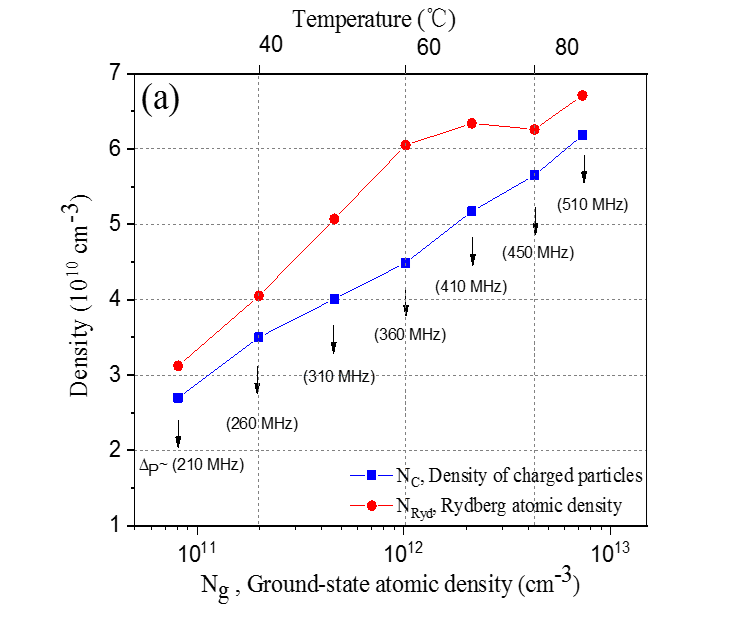}
\end{figure}
\begin{figure}[htbp]
{\includegraphics[width=8.5cm]{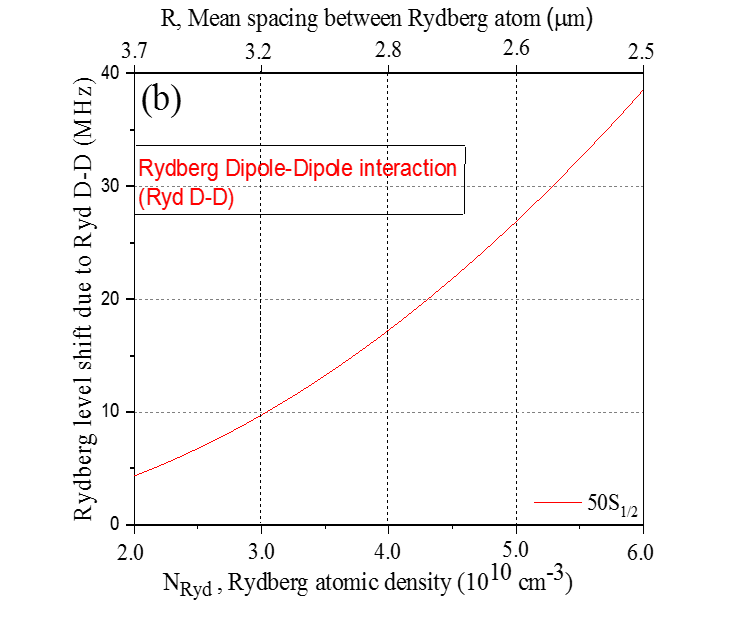}
\includegraphics[width=8.5cm]{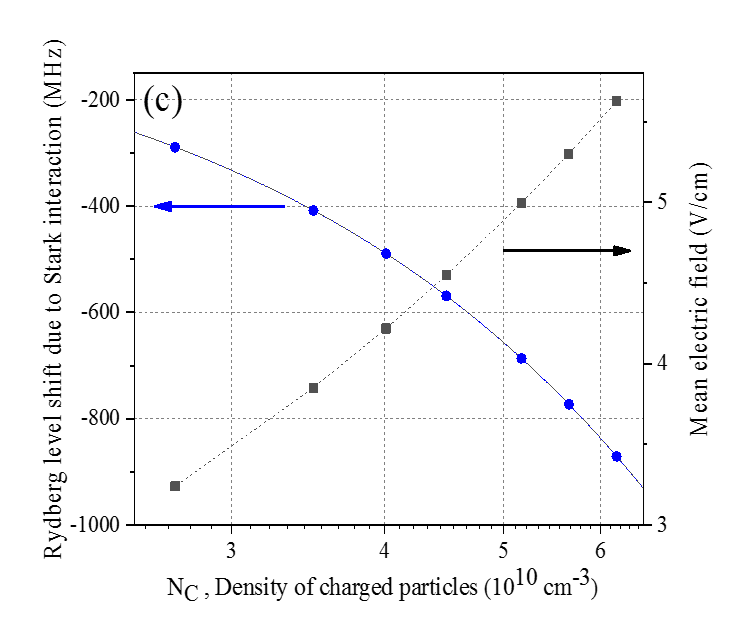}}
\caption{(\textbf{a}) The densities of charged particles and Cs 50S$_{1/2}$Rydberg atoms. The~blue squares indicate the densities of charged particles, while the~red solid circles indicate Rydberg atomic densities; (\textbf{b}) Rydberg level shift due to Rydberg dipole--dipole interaction of Cs atoms for 50$S_{1/2}$ state versus the calculated Rydberg atomic density; (\textbf{c}) Rydberg level shift versus $N_C$ for Cs 50$S_{1/2}$ Rydberg state. The~blue solid circles are Rydberg level shifts caused by the calculated density of charged particles, according to seven temperature conditions of Figure 3(c), while~the blue solid line shows Equation~(3). The~black squares indicate the mean electric fields based on the calculated density of charged~particles.}
\end{figure}

We obtain the densities of charged particles and Rydberg atoms according to the rate equations, as~shown in Figure 4(a). {{It can be seen that the Rydberg atomic density increases first nearly linear with increasing temperature, but~at relatively high temperature, this changes to a slowly increase.}} This may be caused by the different two-step Rydberg excitation (i.e., laser Rabi frequency remains constant, but~the single-photon detuning is quite different). The~single-photon detuning is relatively small under low temperature conditions, so that the two-step Rydberg excitation is relatively strong, which is just the opposite situation under high-temperature conditions. According to Rydberg atomic density, we can calculate Rydberg level shift for the 50$S_{1/2}$ state \cite{ref-journal22}, within a short-distance range, it is a dipole--dipole interaction, which is proportional to 1/$R^3$. The~theoretical result is shown in Figure 4(b). In~the long-distance range, it is Van der Waals interaction, which is proportional to 1/$R^6$. It can be seen that the Rydberg level shift due to dipole--dipole interaction is much smaller than our measured values. Therefore, it is impossible to cause a Rydberg level shift of several hundred MHz. Therefore, the~Rydberg level shift caused by the interaction between Rydberg atoms is not the main factor than our measured values. Here, considering the quantum defect of the Rydberg atoms, we know that there are two types of interaction between the outermost electron of the alkali-metal Rydberg atom and its nucleus: penetration and polarization. As~a result of these two effects, the~atomic binding energy increases and total energy reduces (quantum defect) ~\cite{ref-journal23,ref-journal24}. However, in~general, the~energy loss caused by the quantum defect is very small, so the effect of quantum defect can also be~ignored.

Based on the calculated density of charged particles, the inhomogeneous electric field is \mbox{$\varepsilon=2.603\cdot{\vert e\vert}/(4\pi\varepsilon_0)\cdot N_{C}^{2/3}$} \cite{ref-journal25,ref-journal26}, {{where }}$\varepsilon_0$ is the vacuum dielectric constant, $e$ is the charge of electron.{{ In low electric fields, Stark shift of the Rydberg state $\Delta_S$ is given by second order perturbation theory \cite{ref-journal27} to be $\Delta_S=-(1/2)\alpha\varepsilon^2$,}} where $\alpha$ is the atomic polarizability of Rydberg state. Referring to the related literature ref.\cite{ref-journal28}, the~value for $\alpha$  can be estimated; the polarizability of 50S$_{1/2}$ Rydberg state is $\sim$54.33 MHz·cm$^2$/V$^2$. The~dependence of Rydberg level shift on the density of charged particles can be given as follows \cite{ref-journal25,ref-journal26,ref-journal27}:
\begin{equation}
\Delta_S=-\frac{1}{2}
\alpha(\frac{2.603\cdot{\vert e\vert}}{4\pi\varepsilon_0}\cdot N_{C}^{2/3})^2
\end{equation}

We calculate Rydberg level shift $\Delta_S$ according to Equation~(3), as~shown in Figure 4(c). with~$\Omega_p/{2\pi}$ = 5.0 MHz, $\Omega_c/{2\pi}$ = 36.0 MHz. The blue solid circles in the figure are Rydberg level shifts obtained under seven temperature conditions in Figure 3(c); the~black squares show the calculated mean electric~fields.

Laser beams with different frequencies will interact with atoms of different velocity groups \cite{ref-journal29,ref-journal30}. We can change the frequency detuning of the coupling laser $\Delta_C=-\Delta_P\cdot\lambda_P/\lambda_C$ by controlling the frequency detuning of the probe laser $\Delta_P$, as~the Rydberg level shift can also react to the frequency detuning of the probe laser. Therefore, according to the theoretical simulation, we find $\Delta_S\approx-\Delta_{P,max}\cdot\lambda_P/\lambda_C$, where $\Delta_{P,max}$ is the 852-nm probe detuning of the maximum EIT signal intensity. Therefore, Rydberg level shift $\Delta_S$ is caused by the mean electric field generated by the collisions induced ionization, such that the probe laser {{is in resonance}} with the Cs $6S_{1/2}(F=4)\rightarrow 6P_{3/2}(F'=5)$transition, while the coupling laser {{is in resonance}}  with the Cs $6P_{3/2}(F'=5)\rightarrow 50S_{1/2}$ Rydberg transition, satisfying the two-photon resonance condition, the~intensity of EIT signal is the strongest under this~condition.

We study the relationship between the Rydberg level shift and detuning of the probe laser that produces the EIT signal of maximum intensity multiplied by the Doppler mismatch factor, as~shown in Figure 5.  The blue solid circles indicate the change of Rydberg level shift $\Delta_S$ with density of charged particles, while the red squares indicate the change of probe laser detuning multiplied by the Doppler mismatch factor with density of charged particles, that is, how $-\Delta_{P,max}\cdot\lambda_P/\lambda_C$ varied with density of charged particles. It can be seen that, with the change of density of charged particles, Rydberg level shift and detuning of probe laser multiplied by the Doppler mismatch factor are approximately equal. The~solid black line is the fitting of the change of $-\Delta_{P,max}\cdot\lambda_P/\lambda_C$ with the density of charged particles by using Equation (3). In~our experimental system, the~main deviation is inaccuracy of the temperature control of the atomic vapor cell. For~the red detuning for Rydberg state 50$S_{1/2}$, there is also an energy shift and maximum intensity of Rydberg EIT signal. For~the case of the Cs  $6P_{3/2}(F'=5)\rightarrow 50S_{1/2}$ Rydberg transition, the~red detuning of $\sim$500 MHz is corresponding to the blue detuning of the Cs $6P_{3/2}(F'=3)\rightarrow 50S_{1/2}$Rydberg transition. With~Rydberg EIT spectra, it is difficult to eliminate the influence of hyperfine folds in 6P$_{3/2}$ state. Therefore, the detailed discussion is not performed on the red detuning case for  50$S_{1/2}$Rydberg state. 

\begin{figure}[htbp]
\centering
\includegraphics[width=8cm]{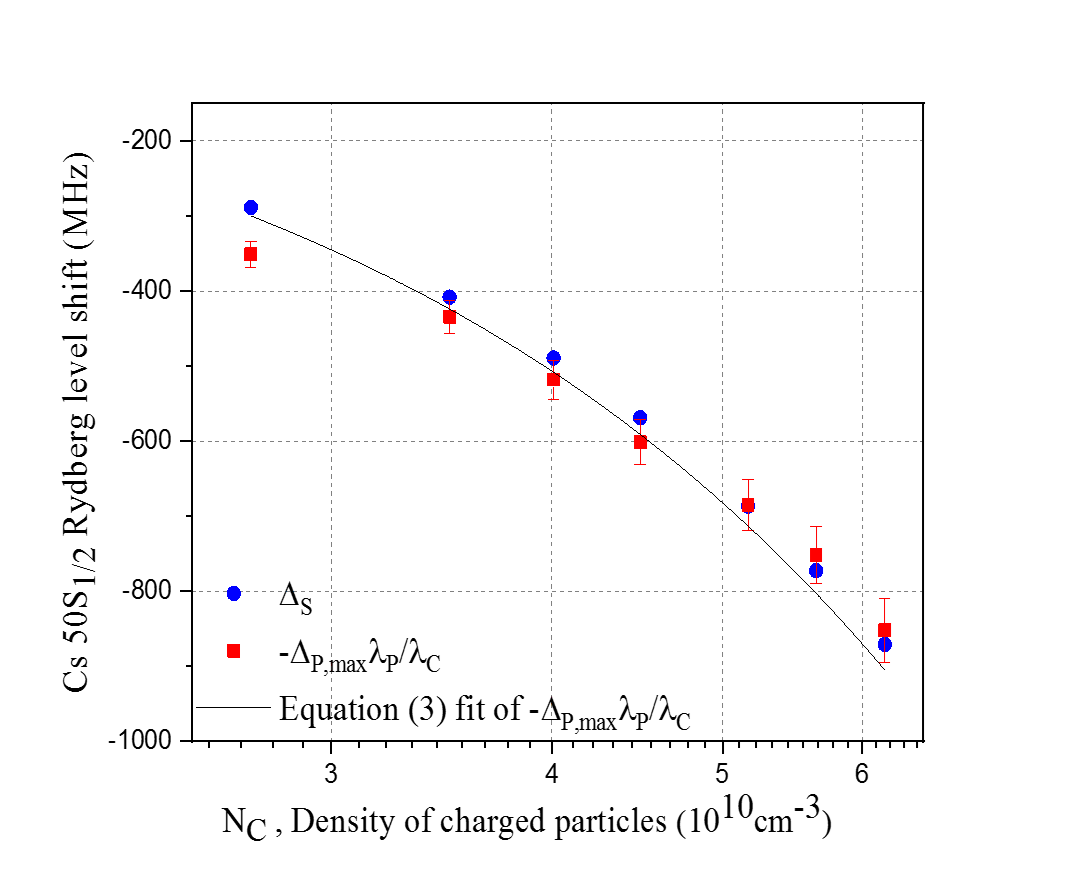}
\caption{Comparison of the relationship between $\Delta_S$ and $-\Delta_{P,max}\cdot\lambda_P/\lambda_C$, where $\Delta_S$ is the calculated Rydberg level shift, $-\Delta_{P,max}\cdot\lambda_P/\lambda_C$ is the detuning of the probe laser that produces the EIT signal of maximum intensity multiplied by the Doppler mismatch factor. The~blue solid circles indicate the calculated Rydberg level shift versus $N_{C}$ for the Cs 50$S_{1/2}$ Rydberg state, the~red squares with error bars indicate the measured probe detuning multiplied by the Doppler mismatch factor versus $N_{C}$ for the Cs 50$S_{1/2}$ Rydberg state. The~solid black line is the fitting of Equation~(3) to $-\Delta_{P,max}\cdot\lambda_P/\lambda_C$.
}
\end{figure}

\section{Conclusions}
In conclusion, {{we experimentally investigated the variation of Rydberg EIT signal intensity upon laser detuning and we found that the intensity of the EIT signal was the highest under a specific detuning of the probe laser,}} when the other experimental conditions were fixed. Then, we analyzed the Rydberg level shift generated by the mean electric field induced by collisions between Rydberg atoms and between Rydberg atoms and ground-state atoms. {{The Rydberg level shift relates to the maximum EIT signal due to the detuning of probe laser.}} Furthermore, we solved the rate equations by using a semi-classical model, and obtained the densities of Rydberg atoms and charged particles generated by Rydberg atoms collision induced ionization. Our results have direct consequences for quantum sensing devices, where Rydberg EIT is employed for sensing radio-frequency and microwave fields.

\begin{acknowledgments}
This research is partially funded by the National Key R$\&$D Program of China (2017YFA03040502), the~National Natural Science Foundation of China (61875111, 11774210, 11974226, and~61905133), the S$\&$T Innovation Programs of Higher Education Institutions in Shanxi Province(2017101), and~the Shanxi Provincial 1331 Project for Key Subjects Construction.
\end{acknowledgments}

\end{document}